\documentstyle[11pt,newpasp,twoside,epsf]{article}
\def\edcomment#1{\iffalse\marginpar{\raggedright\sl#1\/}\else\relax\fi}
\reversemarginpar

\begin{document}
\title{X-ray Luminosity Functions of Young Stars in Taurus-Auriga-Perseus}
 \author{Beate Stelzer}
\affil{Max-Planck-Institut f\"ur extraterrestrische Physik,
 Giessenbachstrasse 1, D-85741 Garching Germany}

\begin{abstract}
A detailed discussion of the X-ray luminosity functions of pre-main
sequence and young main-sequence objects in the Taurus-Auriga-Perseus
region based on the {\em ROSAT} archive is presented. 
One of the main conclusions is that the XLF of classical and weak-line
T Tauri stars (TTS) are different: 
weak-line TTS are X-ray brighter than classical TTS.
Various possible biases related to the sample selection are 
described and ruled out as a reason for the observed discrepancies. 
The X-ray emission of the TTS is compared to that of the Pleiades
and the Hyades to examine the influence of stellar evolution on the
activity level.
\end{abstract}

\section{Introduction}

Within the last years extensive X-ray observations of star forming regions
have established that the youngest among the stellar coronal X-ray
sources are characterized by the strongest activity levels.
A standard picture of stellar magnetic 
activity has developed in which the activity decreases as the stars
evolve from the pre-main sequence (PMS) to the main sequence (MS), and 
continues to decline during the MS life.

While in some early investigations of the X-ray emission of PMS objects 
a power-law relation between age and stellar activity was found, 
other observations
seem to indicate that activity remains relatively constant during the
PMS and starts to drop rapidly once the MS is reached 
(Feigelson \& Kriss 1989, Walter et al. 1988).
Instead of being a pure age effect the decline of activity could be
related to the rotational evolution of the objects: Once on the MS 
the stars spin down and the dynamo should become less efficient.
Many observational studies concerning the connection between 
rotation and X-ray activity 
have indicated a correlation between 
X-ray emission and rotational velocity favoring the spin-down scenario
(see e.g. Bouvier 1990; Neuh\"auser et al. 1995 = N95; Stauffer et al. 1997). 
However, these results were contradicted by other studies 
(Gagn\'e et al. 1995; Micela et al. 1996).
Thus, despite all efforts, the details of the relation between 
age, activity, and rotation are still not well understood. 

An important tool to assess the strength of the X-ray activity 
are X-ray Luminosity Functions (XLF), i.e. cumulative distributions of
X-ray luminosities $L_{\rm x}$. Previous analyses of XLF of young stars have
either focused on individual X-ray exposures of selected sky regions or 
were based on spatially extended but low-sensitivity observations,
such as e.g. the {\em ROSAT} All-Sky Survey (RASS). 

In this contribution a systematic analysis of the XLF of young stars
in Taurus-Auriga-Perseus on basis of 
all publicly available {\em ROSAT} PSPC observations pointed towards that
portion of the sky is presented.
The stellar sample includes
both PMS objects (T Tauri Stars from Taurus-Auriga), and 
representatives of young MS stars (the Pleiades and the Hyades cluster).
A total of 106 observations are evaluated and more than 800 sources detected.
The statistical analysis was performed with the ASURV package 
(Feigelson et al. 1985) to take proper account of upper limits for
non-detections.

\section{XLF of samples with unresolved binary stars}

Most of the binary stars in the region under study are unresolved with
the {\em ROSAT} PSPC (spatial resolution $\sim 25^{\prime\prime}$ on-axis).
The emission from unresolved multiples is difficult to 
interpret because it is not clear which of the components plays the active 
part. In a study of TTS binaries 
in the Taurus-Auriga region which are resolved 
by the {\em ROSAT} HRI K\"onig et al. (2000) found that both components
emit X-rays. Therefore, it seems adequate to split the observed $L_{\rm x}$ 
in equal amounts on the number of components in each (unresolved) 
stellar system. As a check of this hypothesis separate XLF for single and
binary stars have been computed for each sample (i.e. TTS, Pleiades
and Hyades). Because of the known dependence of X-ray emission on 
spectral type we consider G, K, and M stars separately.
Binary stars are treated in two ways: (A) according to the assumption 
described above, and (B) without taking account of the multiplicity,
i.e. by attributing all observed X-rays to the primary in the system.
Fig.~1 visualizes the effects this different assumptions have 
on the XLF. Distributions of binaries treated as in (A) are termed `b2',
and those of type (B) are termed `b1'. In Fig.~1 the XLF of both types
of binary distributions is compared to that of single stars.
As expected the distributions `b1' lie off to 
the right of the distributions `b2'. Statistical tests reveal that the
latter ones agree with the distributions of singles, but the `b1' 
distributions are significantly different. 

Pye et al. (1994) and Stern et al. (1995) have discussed the XLF of
Hyades stars based on a smaller set of {\em ROSAT} observations. They 
found that among the Hyades stars of spectral type K the binaries are
stronger X-ray emitters than the single stars. The distributions
derived in these earlier studies correspond to our type `b1' distribution.
This indicates that a proper
treatment of the binary character is important in understanding the
XLF if the sample includes unresolved multiples.
In view of the above result all XLF discussed furtheron in this paper
are based on choice (A).
\begin{figure}[ht]
\plotfiddle{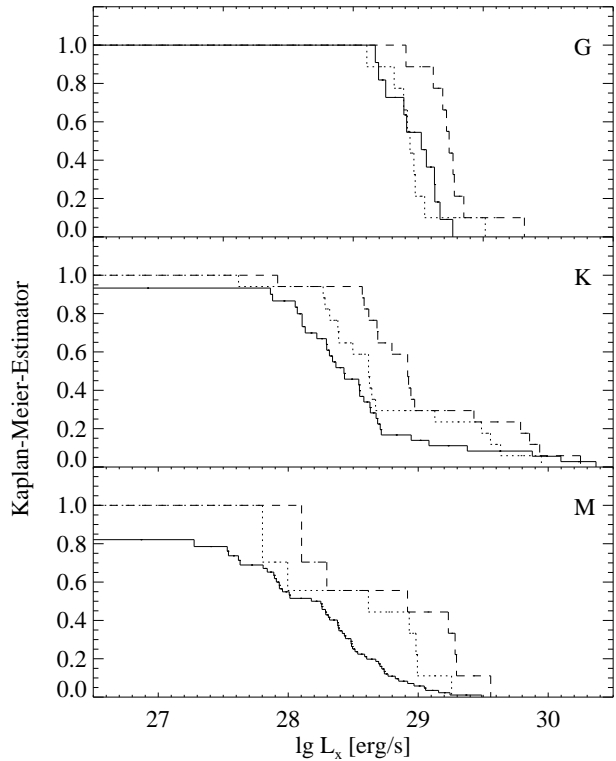}{10.4cm}{0}{54}{54}{-120}{0}
\caption{XLF of single and binary stars in the Hyades. {\em solid} - single
stars, {\em dotted} - binaries `b2', and {\em dashed} - binaries `b1' (see
text for a description of the samples). Separate XLF have
been compiled for G, K, and M stars, and demonstrate the trend towards
lower $L_{\rm x}$ for cooler stars.}
\end{figure}

\section{XLF of cTTS and wTTS}

In an analysis of RASS observations in Taurus-Auriga Neuh\"auser et al. (1995)
found that cTTS and wTTS are statistically different concerning the
amount of X-rays emitted: wTTS show stronger X-ray emission than cTTS.
This can in principle 
be explained in terms of rotational evolution, because the wTTS -- 
having lost their disks -- can spin up and drive a more efficient dynamo,
while for the cTTS the rotation rate is restrained by magnetic coupling
between the star and its accretion disk. Contrary to this obvious
explanation, X-ray
studies in other star forming regions (Feigelson et al. 1993; Grosso et
al. 2000) do not show differences between the activity of cTTS and wTTS.

In order to test the XLF for various contaminating effects 
a systematic analysis of pointed {\em ROSAT} observations
in the Taurus-Auriga region (including the Pleiades and Hyades)
is performed. The resulting XLF for cTTS and wTTS are displayed in 
Fig.~2. The distributions derived from this study of pointed PSPC
observations are similar to those obtained earlier by N95 from the
RASS at the high luminosity end, 
but demonstrate the better sensitivity of the pointed observations
in the low-luminosity regime. The wTTS are found to be stronger X-ray
emitters than the cTTS, confirming the result of N95. The remainder of
this paper will be focused on the discussion of several effects that could
be responsible for the observed differences.
\begin{figure}[ht]
\plotfiddle{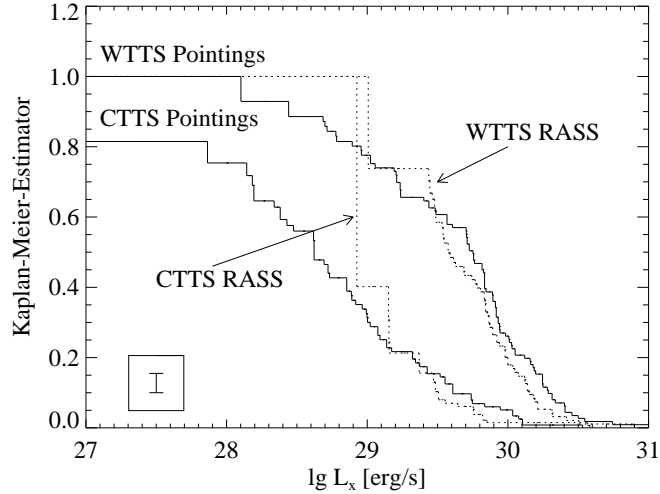}{7cm}{0}{55}{55}{-120}{0}
\caption{XLF of cTTS and wTTS in Taurus-Auriga as derived from the RASS
(dotted lines) and from pointed PSPC observations (solid lines).}
\end{figure}

\subsection{An X-ray selection bias in the sample of wTTS ?}

One argument that can be put forth against results of the type shown in 
Fig.~2 is that the wTTS sample may be biased towards strong X-ray emitters,
because the main identification method for wTTS are indeed their
X-rays. cTTS are not affected by this bias because most of them
have been identified by optical surveys (e.g. H$\alpha$). A possible
selection bias in the wTTS sample is examined by computing
separate XLF for wTTS discovered by X-ray satellites ({\em Einstein} 
and {\em ROSAT}) and those discovered by other methods, 
e.g. proper motion studies. The comparison of
these distributions (displayed in Fig.~3 on the left) shows that there is no
significant difference. This leads to the conclusion that the total sample 
of wTTS is not biased towards strong X-ray emitters.
\begin{figure}
\plottwo{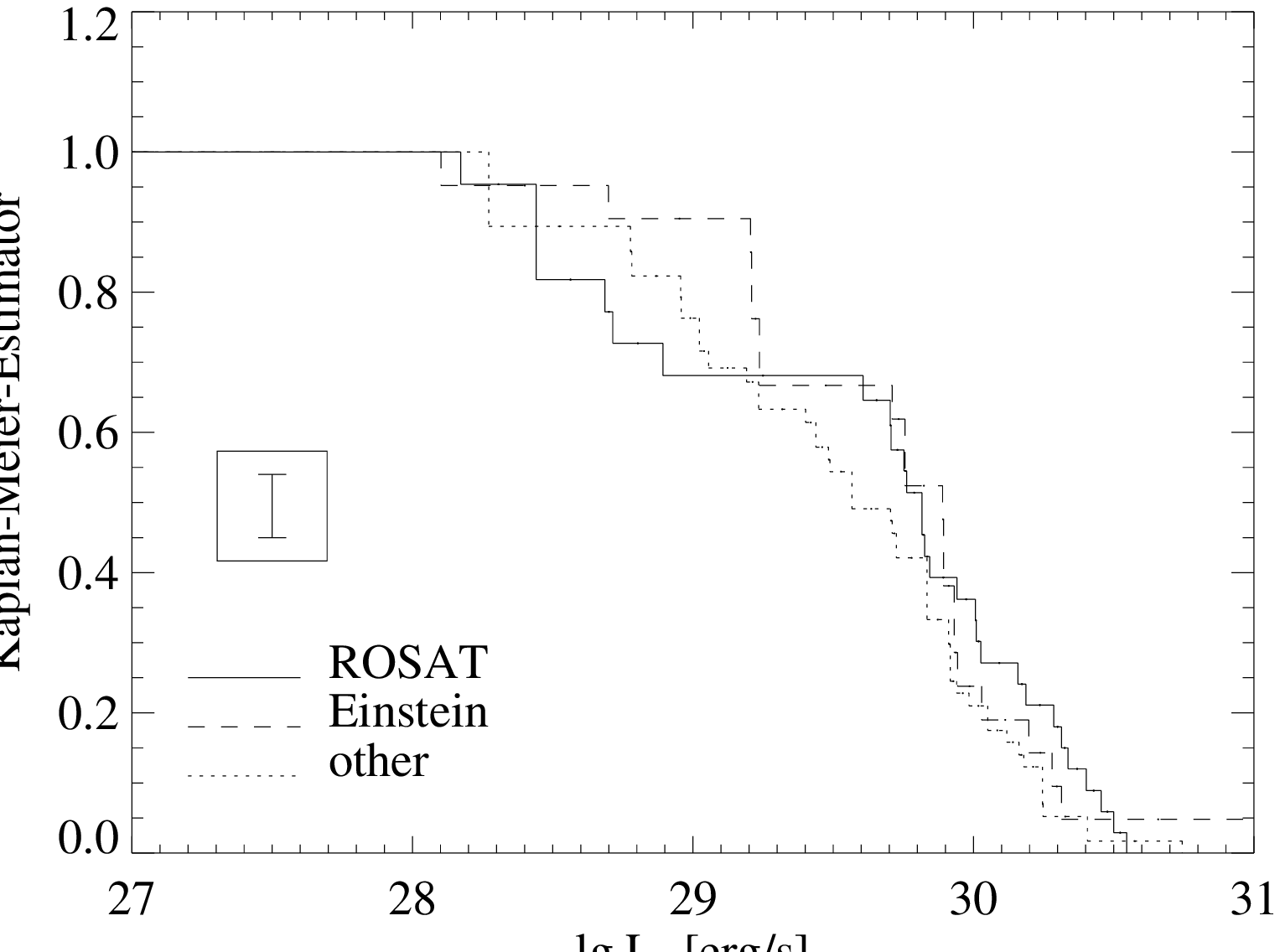}{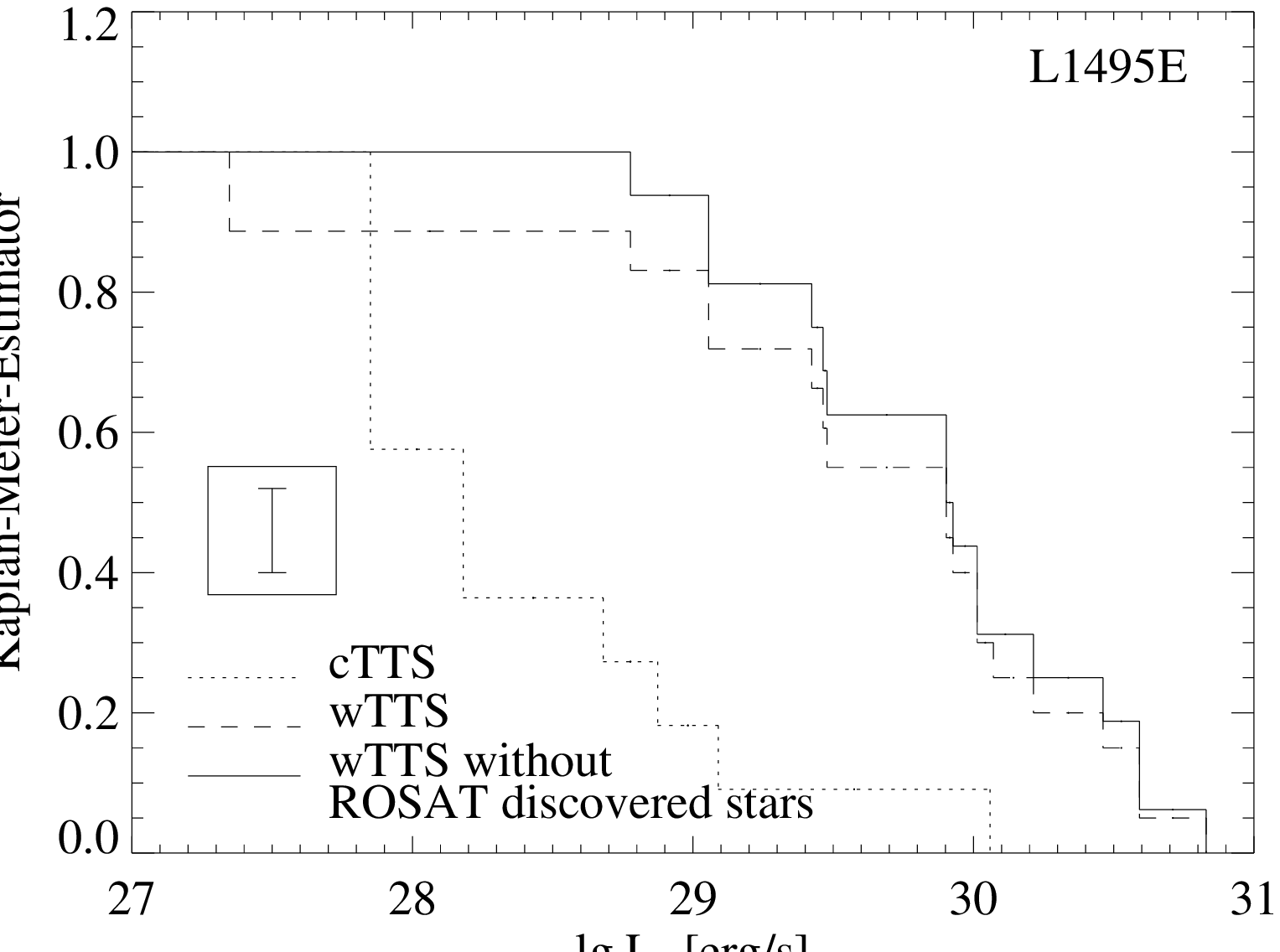}
\caption{{\em left} - XLF for subsamples of wTTS in Taurus-Auriga: {\em
ROSAT} discovered wTTS, {\em Einstein} discovered wTTS, and wTTS discovered
by other means, i.e. not by X-ray emission. {\em right} - XLF of cTTS and
wTTS in L1495\,E, a cloud in the central part of the Taurus star forming complex.}
\end{figure}

\subsection{Age segregation}

A large fraction of the TTS discovered in the RASS are wTTS at the outer
edge of well-known star forming regions. 
Although the individual ages of the stars are subject to large
uncertainties, 
there is a tendency of these objects to be somewhat older than stars
in the central parts of the molecular clouds. In this sense the differences
between the XLF of cTTS and wTTS could be the result of comparing
samples with different ages. XLF from wTTS on the clouds should then be
similar to the XLF of cTTS (most of which are on the clouds). 
Fig.~3 on the right shows a comparison of the XLF of cTTS and
wTTS in the L1594\,E cloud, the region of densest concentration of
molecular material in the Taurus-Auriga complex. The data are taken from 
a long pointed observation ($ROR$ 200001-0 and 200001-1).
The X-ray luminosities of the on-cloud cTTS are lower than those of the 
on-cloud wTTS ruling out 
that the observed difference in the total sample is an 
effect of age segregation.

\subsection{Distribution of spectral types}

The X-ray luminosity of active stars scales with spectral type. This
implies that differing spectral type distributions of cTTS and wTTS 
may have an effect on the XLF. 
However, XLF computed separately for three spectral
type bins show that from G to M stars wTTS are stronger
X-ray emitters than cTTS. Fig.~4 shows the XLF of K and M TTS 
together with Pleiades and Hyades stars of the
same spectral type, and demonstrate the continual 
decline of $L_{\rm x}$ for stars on the MS. The corresponding XLF for G 
stars (not shown) have similar shape but low statistics in the sample of cTTS.
\begin{figure}
\plottwo{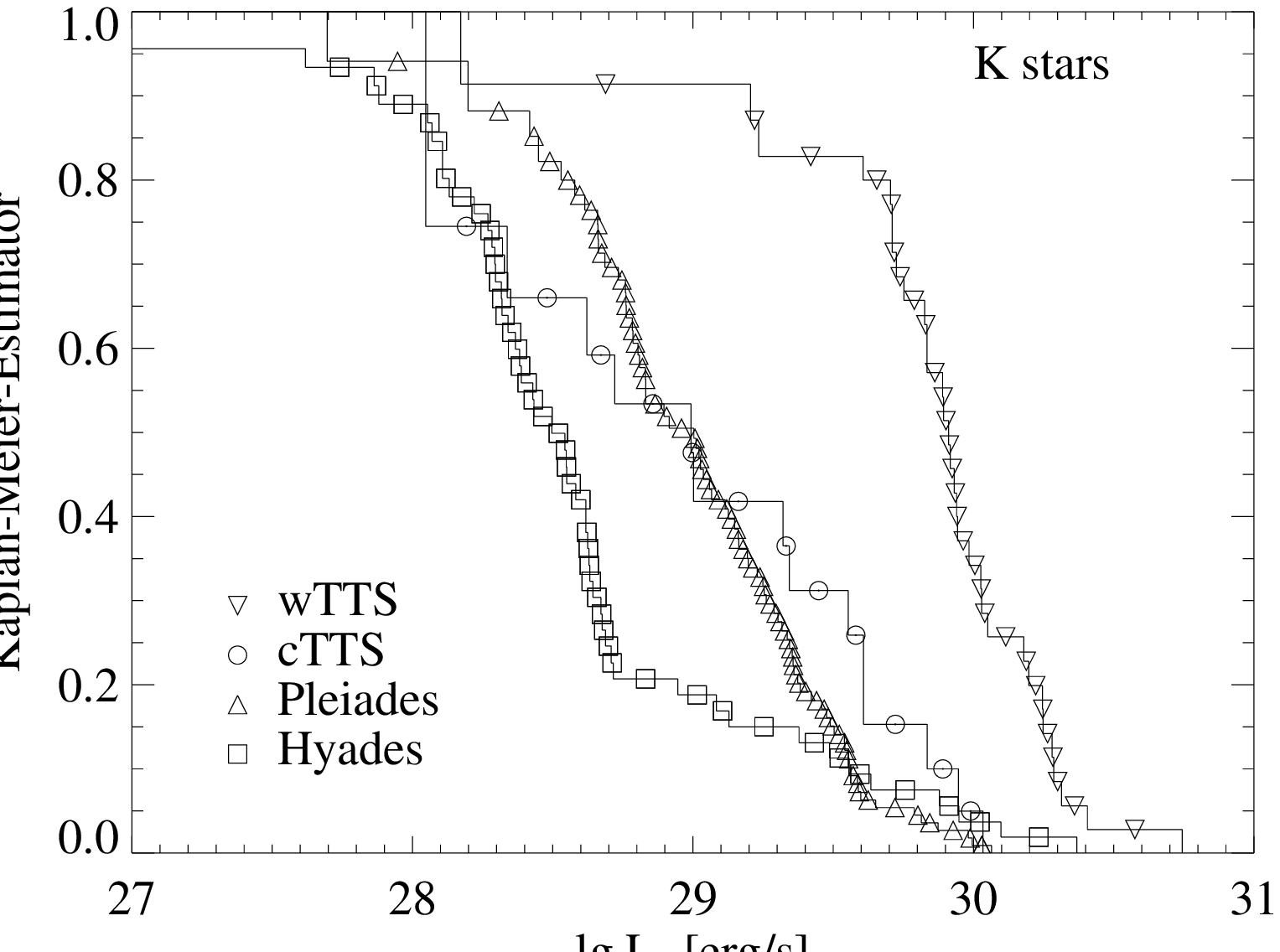}{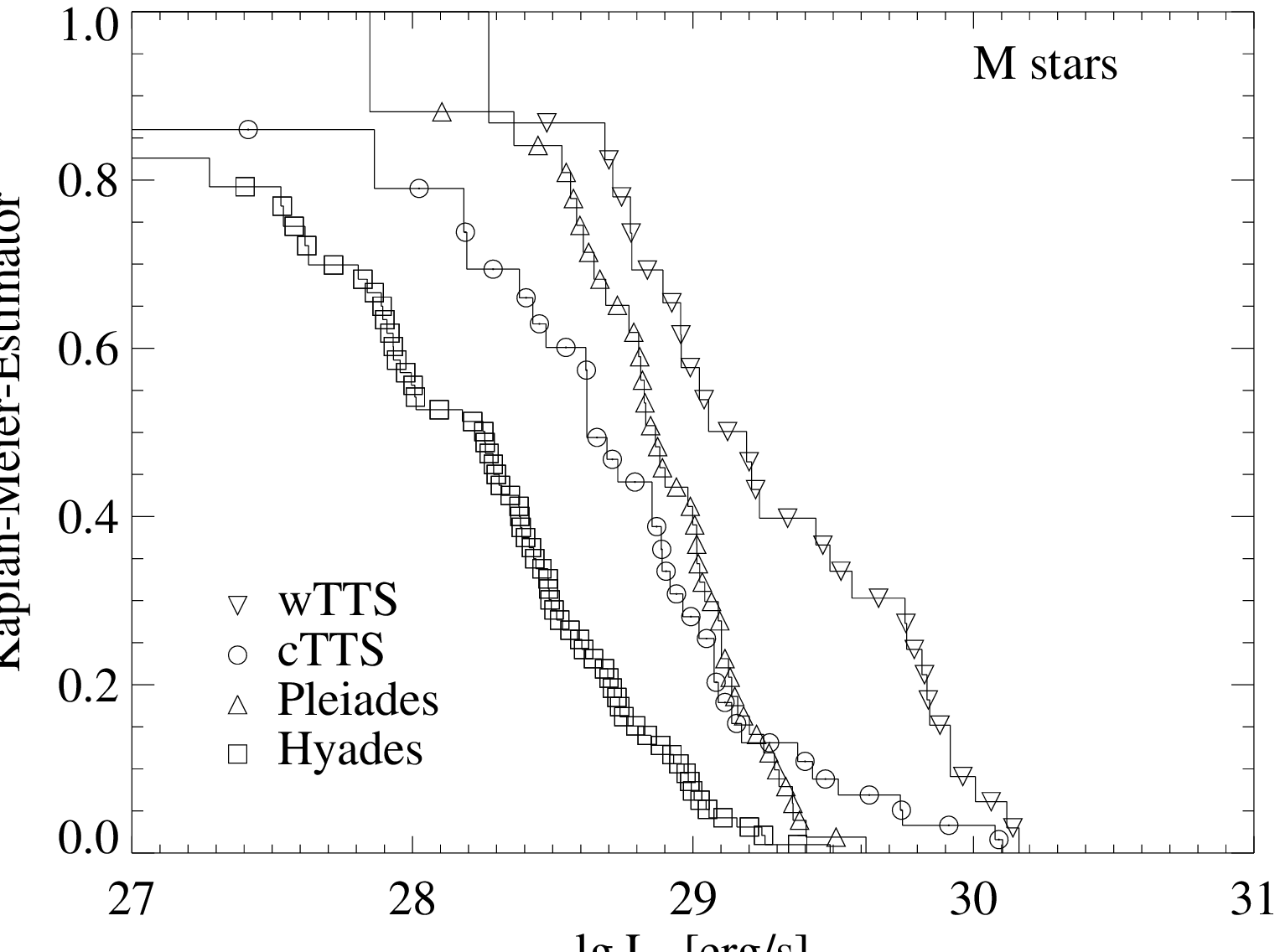}
\caption{XLF for cTTS, wTTS, Pleiades, and Hyades of spectral type
K and M.}
\end{figure}

\section{Conclusions}

A study of the XLF from a large sample of PMS and ZAMS objects was
performed on basis of pointed {\em ROSAT} PSPC observations.

This analysis shows that the XLF are sensitive to the way in which
unresolved binaries are dealt with. Two cases have been considered,
in which the observed luminosity is either all attributed to the
primary or distributed equally onto
all components in multiples. A comparison shows that the latter assumption
is consistent with the XLF of single stars. 
If all X-rays are assumed to come from the primary instead, 
the XLF of binaries are too bright. 

In Taurus-Auriga the wTTS show higher $L_{\rm x}$ than the cTTS
in agreement with earlier studies in the same star forming 
region, but in contrast to results from other regions (e.g. Cha~I, $\rho$~Oph).
A detailed analysis was undertaken to examine possible explanations
for the discrepancy of the XLF of cTTS and wTTS in Taurus-Auriga.
Effects which could be ruled out are
\begin{itemize}
\item an X-ray bias in the wTTS sample,
\item different age distributions of cTTS and wTTS,
\item different spectral type distributions of cTTS and wTTS,
\item uncertainties in the X-ray emission from unresolved binaries (not
discussed in the text).
\end{itemize}
The only obvious parameter left with a possible influence on the X-ray emission
is rotation. A comparison of slow to fast rotating wTTS in Taurus-Auriga
indicates that fast rotators show higher $L_{\rm x}$. However, the sample
of stars with measured rotation periods is small (16 fast versus 12 slow
rotators among the wTTS). 

For a more detailed discussion of the XLF and the rotation-activity
relation we refer to Stelzer \& Neuh\"auser (2000).

\acknowledgements{The {\em ROSAT} project is supported by the
Max-Planck-Gesellschaft and Germany's federal government (BMBF/DLR).}

\end{document}